# Quantum nature of laser light

DAVID T. PEGG

Centre for Quantum Computer Technology, School of Science, Griffith University, Nathan, Brisbane 4111, Australia

JOHN JEFFERS

Department of Physics, University of Strathclyde, Glasgow G4 0NG, U.K.

Author for correspondence: D. T. Pegg

Phone +61 7 3875 7152. Email: D.Pegg@griffith.edu.au

**Abstract.** All compositions of a mixed state density operator are equivalent for the prediction of the probabilities of future outcomes of measurements. For retrodiction, however, this is not the case. The retrodictive formalism of quantum mechanics provides a criterion for deciding that some compositions are fictional. Fictional compositions do not contain preparation device operators, that is, operators corresponding to states that could have been prepared. We apply this to Mølmer's controversial conjecture that optical coherences in laser light are a fiction and find agreement with his conjecture. We generalise Mølmer's derivation of the interference between two lasers to avoid the use of any fictional states. We also examine another possible method for discriminating between coherent states and photon number states in laser light and find that it does not work, with the equivalence for prediction saved by entanglement.

## 1. Introduction

It is usually assumed that the light from a single-mode laser is in a coherent state with a definite but unknown phase. This description can be used successfully for most practical purposes. Such a state gives photocount statistics that are in accord with experiment and has coherence properties similar to those of a classical field, which are useful for explaining interference effects. A coherent state is a superposition of photon number states and its density matrix in the photon number basis has non-zero off-diagonal elements, which we can refer to as optical coherences. As the phase of the laser light at any particular time is considered to be unknown, we can assign to each value of phase an equal *a priori* probability of occurring. This can be expressed by writing an ensemble averaged density operator for the laser light as

$$\hat{\rho}_F = \int_0^{2\pi} |\alpha\rangle\langle\alpha| \frac{d\theta}{2\pi} \tag{1}$$

where $|\alpha\rangle$, with $\alpha = |\alpha|\exp(i\theta)$, is a coherent state. This is in accord with standard derivations of the density operator for the laser field.

In an important and controversial paper, Mølmer [1] has questioned whether the standard interpretation above is correct. He conjectures that the optical coherences are merely a convenient fiction. The density operator in (1) can be partitioned in many different ways, including

$$\hat{\rho}_F = \exp(-|\alpha|^2) \sum_{n=0}^{\infty} \frac{|\alpha|^{2n}}{n!} |n\rangle\langle n| \quad . \tag{2}$$

Expressions (1) and (2) produce the same diagonal density matrix in the number state representation with the off-diagonal elements, or coherences, averaging to zero if we start with (1), or being identically zero in each term if we take the matrix elements of (2). Just as (1) is what we would write if we knew the laser field was prepared in a coherent state, but have no information about its phase, expression (2) is what would write if we knew that the field was prepared in a number state but only have probabilistic information as to which one. The actual density operator, that is what we would write if we had sufficient information about the preparation, would possess coherences in the first case but not in the second case. An immediately obvious objection to the second interpretation is that interference between beams from two lasers has been observed [2]. This seems to support the idea that laser light has a well-defined, if unknown, phase and thus supports the coherent state interpretation. Mølmer [1, 3], however, has examined the interference between two beams from laser cavities that are initially in identical number states and shows, surprisingly, that similar interference effects are obtained.

This is an example of the indistinguishability of different partitions of a density operator by means of measurement outcomes [4, 5]. Preferring one partition to another is sometimes called the "preferred ensemble fallacy" or "partition ensemble fallacy" [6-8]. Essentially $\hat{\rho}_F$ itself does not contain enough information to enable us to say what the real state is. Furthermore we cannot distinguish different partitions of $\hat{\rho}_F$ by subsequent measurement, because the probability of the outcome $j$ of any measurement of the light is given by $\text{Tr}(\hat{\rho}_F \hat{\Pi}_j)$, where $\hat{\Pi}_j$ is the associated element of a probability operator measure (POM), and the measurable expectation value of any observable $\hat{A}$ is given by $\text{Tr}(\hat{\rho}_F \hat{A})$. Neither $\hat{\Pi}_j$ nor $\hat{A}$ add any information about the state itself. It follows that all statistical *predictions* for the ensemble are determined by $\hat{\rho}_F$ and not by its real composition [4]. While a density operator can be constructed from knowledge of the component states and their corresponding *a priori* probabilities, it is not possible to reverse the procedure and infer the composition of the mixture uniquely from the density operator itself.

Mølmer's conjecture with its immediate ramifications, for example that squeezed light as presently generated is also a myth, has certainly not been universally accepted and has been argued against [8-10]. Most physicists still prefer to accept that lasers produce coherent states. The issue was revived by Rudolph and Sanders [7] in respect to continuous variable quantum teleportation [7]. They argue that the appearance of success of experiments on this type of teleportation relies on interpreting the real state of the laser light used as being a coherent state of unknown phase, in disagreement with Mølmer's conjecture. The arguments of Rudolph and Sanders have also not received widespread acceptance, with continuous variable quantum teleportation being defended [8, 11].

The preferred ensemble fallacy raises the question as to whether or not the reality of a particular composition of the density operator has any meaning. For *predictive* purposes we have freedom to choose any composition we like. Does this mean that we are at liberty to say that any composition is as real as any other and thus it is a matter of choice whether we say the experiments do or do not demonstrate continuous variable quantum teleportation? In this paper, we examine this question. The real state of a system is correlated with the state in which the system was actually prepared. Inferring the actual state in which a system was prepared from a knowledge of measurement outcomes is a matter of *retrodiction* [12, 13]. The retrodictive formalism of quantum mechanics [13] gives a means of doing this. Just as there are measurement device operators, or POM elements, that contain information about how a system will be measured, in the retrodictive formalism there are preparation device operators that contain information about the way in which a system was prepared. We find that the retrodictive formalism does allow us to ascertain preferred compositions and thus attach notions of reality and fiction to particular states. We apply this in the context of examining the state of laser light. A difficulty with laser light is that describing it by any unentangled density operator such as $\hat{\rho}_F$ is most likely a fiction in itself. This leads us to generalise Mølmer's mechanism for the interference between two lasers to avoid invoking any fictional states.

**2. Symmetric quantum mechanics and causality**

Because of the unfamiliarity of the quantum retrodictive formalism we give a brief outline here. Consider an experiment where Alice prepares a system in some state which we associate with a preparation event *i* and then, before the system has had time to evolve significantly, Bob performs a measurement on the system with an outcome *j*. This experiment is repeated many times and list of combined events (*i*, *j*) for each experiment is constructed. The basic postulate connecting quantum mechanics to probability can be expressed in the preparation-measurement symmetric form [13]

$$P(i,j) = \frac{\mathrm{Tr}(\hat{\Lambda}_i \hat{\Gamma}_j)}{\mathrm{Tr}(\hat{\Lambda}\hat{\Gamma})} \qquad (3)$$

where $P(i,j)$ is the probability for the combined event as measured by the occurrence frequency on the list, $\hat{\Lambda}_i$ and $\hat{\Gamma}_j$ are positive or negative definite operators acting on the Hilbert space of the system, $\hat{\Lambda} = \sum_i \hat{\Lambda}_i$ and $\hat{\Gamma} = \sum_j \hat{\Gamma}_j$. The set of $\hat{\Lambda}_i$ ($\hat{\Gamma}_j$), which are called preparation (measurement) device operators, provides a mathematical description of the action of the preparation (measurement) device. As multiplication of $\hat{\Lambda}_i$ or $\hat{\Gamma}_j$ by a constant does not alter $P(i,j)$ we can, without loss of generality, set $\text{Tr}\hat{\Lambda} = 1$ and $\text{Tr}\hat{\Gamma} = 1$. Let us assume that both Alice and Bob faithfully record every preparation event and associated measurement event respectively. Then we can find the probability, or occurrence frequency, for preparation event $i$ by summing $P(i,j)$ over $j$. This yields

$$P(i) = \frac{\text{Tr}(\hat{\Lambda}_i \hat{\Gamma})}{\text{Tr}(\hat{\Lambda}\hat{\Gamma})} \ . \tag{4}$$

Suppose the series of Alice's preparation events on identical systems takes an hour and Bob performs the corresponding measurements the next day with his choice of measuring device. The occurrence frequency $P(i)$ can be easily ascertained by Alice as soon as she has finished the series of preparation events. From (4), however, we see that $P(i)$ is a function of $\hat{\Gamma}$, which relates to Bob's measuring apparatus. If Bob were to have some control over $\hat{\Gamma}$, he could use this control to affect $P(i)$ in (4) and thus send a message to Alice which she would receive on the previous day, which would violate causality. To preserve causality, we must therefore ensure that Bob's choice of measuring device cannot affect $P(i)$ given by (4). If Bob is using a Stern-Gerlach apparatus, say, then, by selecting any orientation of the magnetic field at will, he can perform a unitary transformation on $\hat{\Gamma}$. If this were to change $\hat{\Gamma}$ appropriately, he could exert some control over $P(i)$. To prevent this, that is to impose causality, we must demand that $\hat{\Gamma}$ be unaffected by any unitary transformation and thus must be proportional to the unit operator, that is $\hat{\Gamma} = k\hat{1}$. Then $P(i) = \text{Tr}\hat{\Lambda}_i$, which, because this is determined solely by Alice, we can call the *a priori* probability of preparing state $i$. Defining $\hat{\Pi}_j \equiv \hat{\Gamma}_j / k$, we see that $\hat{\Pi}_j$ are positive definite operators that sum to the unit operator and are thus the elements of a probability operator measure (POM). We can also define $\hat{\rho}_i \equiv \hat{\Lambda}_i / \mathrm{Tr}\hat{\Lambda}_i$, which will be a positive definite operator with trace unity, that is, a density operator. Then we can write the probability of detection event $j$ given preparation event $i$ as

$$P(j|i) = P(i,j)/P(i) = \mathrm{Tr}(\rho_i \hat{\Pi}_j), \tag{5}$$

which is the standard postulate for predictive quantum mechanics. From the above we see that this standard expression incorporates causality.

The corresponding retrodictive expression $P(i|j)$, that is, the probability that Alice prepared state $i$ if Bob's measurement event is $j$, is given by

$$P(i|j) = P(i,j)/P(j) = \frac{\mathrm{Tr}(\hat{\Lambda}_i \hat{\Pi}_j)}{\mathrm{Tr}(\hat{\Lambda} \hat{\Pi}_j)}. \tag{6}$$

We can perform this retrodictive calculation if we know the set of preparation device operators $\hat{\Lambda}_i$, that is, the action of the preparation device. Noting from above that $\hat{\Lambda}_i = P(i)\hat{\rho}_i$, we see that the retrodiction is possible if we know the states that Alice can possibly prepare and the *a priori* probabilities of her doing so. Further, because $\hat{\Lambda} = \sum_i \hat{\Lambda}_i$, we see that $\hat{\Lambda}$ is just the density operator, $\hat{\rho}$ say, that we would assign to the state if we knew the set of preparation device operators $\hat{\Lambda}_i$ but do not know which individual state was prepared and have no measurement information. To shorten the discussion, we have left out time evolution here but this can be incorporated. For a closed system we can apply either a unitary forward-time evolution operator to the preparation device operators or a unitary backward-time evolution operator to the measurement POM elements [13]. The cyclic property of the trace ensures that the same probabilities are obtained. For open systems the situation is more complicated but is tractable [14].

From the above we see that we can assign a density operator to the prepared state, in the absence of knowledge of the actual state prepared or any measurement information, given by

$$\hat{\rho} = \sum_i P(i)\hat{\rho}_i \qquad (7)$$

where the sum is over the states that can possibly be prepared. Subsequent information about the outcome of a measurement allows us to find the probability that a particular state was actually prepared. From (6) and (7) this is

$$P(i|j) = \frac{\text{Tr}[P(i)\hat{\rho}_i\hat{\Pi}_j]}{\text{Tr}(\hat{\rho}\hat{\Pi}_j)} \; . \qquad (8)$$

Comparing (8) with (5), we note that (5) is predictive, giving the probability that a particular measurement event will take place given the outcome of a preparation event; (8) is retrodictive, giving the probability that a particular preparation event did take place given the outcome of a measurement event. The lack of symmetry in form arises from the different normalisation conditions imposed on the preparation and measurement device operators by our requirement of causality [15]. Later we shall apply (8) to study the case of the interference of light leaking from two cavities containing fields prepared in coherent states.

## 3. Retrodiction and the preferred ensemble fallacy

We now examine the preferred ensemble fallacy. If the best description we can give to the prepared state is $\hat{\rho}$, then statistical predictions of future outcomes are determined by (5) with $\hat{\rho}_i$ replaced by $\hat{\rho}$ and where $\hat{\Pi}_j$ is the POM element evolved back in time to the time of preparation. Although we can also express $\hat{\rho}$ mathematically as linear combinations of states other than those in (7), some of which may be impossible to prepare with the particular preparation device, this will not affect these predictions. For making retrodictions, on the other hand, expansion (7) most certainly is a preferred decomposition. Other expansions will not give the correct preparation device operators that are essential for use in the numerator of (6) or (8) allowing us to calculate the correct retrodictive probabilities that particular states were actually prepared. The simplest illustration of this is as follows. Suppose Alice has, for example, a Stern-Gerlach apparatus that prepares a spin-half atom in the states $|+z\rangle$ or $|-z\rangle$ and she prepares these with equal probability. The preparation device operators that describe this device mathematically are $\hat{\Lambda}_+ = |+z\rangle\langle+z|/2$ and $\hat{\Lambda}_- = |-z\rangle\langle-z|/2$. Bob has a measurement device described mathematically by the measurement device operators $\hat{\Pi}_+ = |+x\rangle\langle+x|$ and $\hat{\Pi}_- = |-x\rangle\langle-x|$ (which are elements of a probability operator measure). In the absence of knowledge as to which state Alice prepared, $\hat{\rho} = \hat{1}/2$ and the probability of a measurement outcome $+x$ is 1/2. The probability of a measurement outcome $-x$ is also 1/2. The probability of any other state, such as $|+y\rangle$, being measured is zero as $|+y\rangle\langle+y|$ is not a measurement device operator. In the retrodictive case, we find from (6) that if the measurement outcome is $+x$ or $-x$ then the probability that the prepared state was $|+z\rangle$ is 1/2 as is the probability that the prepared state was $|-z\rangle$. The probability that the prepared state was $|+y\rangle$, say, is zero as $|+y\rangle\langle+y|/2$ is not a preparation device operator. This is true even though we can write $\hat{\rho} = \hat{1}/2$ as $|+y\rangle\langle+y|/2 + |-y\rangle\langle-y|/2$. Thus the composition of the density operator in terms of the actual preparation device operators is essential for inferring what state was actually prepared. In only this composition are the coefficients preparation probabilities. There is certainly a preferred ensemble.

## 4. Preparation of coherent states

As the measured statistical properties of laser light are determined by a density operator $\hat{\rho}_F$, the correlation functions indicating various orders of coherence in the sense used by Glauber [16], for example in terms of maximum fringe contrast, will be unaffected by a particular composition. Only in situations such as that raised by Rudolph and Sanders in respect to continuous variable quantum teleportation [7], where a decision must be made whether or not laser light really is in a coherent state with unknown phase, is the composition important. From our preceding discussion, the state that a laser is really in is determined by the state in which it is prepared. This in turn is determined by the preparation device operators which themselves are determined by the physical action of the preparation device. Standard theories of the laser [17], unfortunately, aim only at determining $\hat{\rho}_F$ itself. Mølmer [1] has briefly discussed some physical reasons for his conjecture that the coherent state description of laser light may be wrong. In this section, we look at this question a little more closely.

Often a coherent state is regarded as that produced by a classical current such as an oscillating charged body. This is because mathematically we can create a coherent state from the vacuum by displacing it with a Glauber operator $\exp(\alpha \hat{a}^\dagger - \alpha^* \hat{a})$ [18] where the operators are creation and annihilation operators and $\alpha$ is a c-number. This displacement operator corresponds to a unitary evolution operator for a classically oscillating source coupled to a field mode of the same frequency. It is worth examining firstly, therefore, the broader question as to whether or not it is indeed possible to produce a coherent state from a quantum mechanical source. Let the field mode with frequency $\omega$ be coupled to a quantum source system with gaps between energy levels of $\hbar\omega$. Let the interaction Hamiltonian be of an energy-conserving form

$$\hat{H}_I = i\lambda(\hat{a}^\dagger \hat{c} - \hat{c}^\dagger \hat{a}) \tag{9}$$

that commutes with the total free Hamiltonian of the source plus the field. The operator $\hat{c}$ acts on the state space of the source and $\hat{a}$ is the annihilation operator of the field. Then in the interaction picture the Hamiltonian will be just $\hat{H}_I$. In the Heisenberg representation we easily obtain from (9)

$$\frac{d\hat{a}(t)}{dt} = \lambda \hat{c}(t) \; . \tag{10}$$

Let the initial state of the combined system be $|\gamma\rangle_S |0\rangle_F$. The vacuum state $|0\rangle_F$ of the field is a coherent state, that is an eigenstate of $\hat{a}$, with complex amplitude $\alpha(0) = 0$. We want the field to stay in an eigenstate of $\hat{a}$ at later times $t$ but with a non-zero amplitude $\alpha(t)$. That is we want

$$\hat{a}\hat{U}(t)|\gamma\rangle_S |0\rangle_F = \alpha(t)\hat{U}(t)|\gamma\rangle_S |0\rangle_F \tag{11}$$

where $\hat{U}(t)$ is the time displacement operator for the Hamiltonian (9). If (11) is true then, by letting $\hat{a}(t) = \hat{U}^{-1}(t)\hat{a}\hat{U}(t)$ act on $|\gamma\rangle_S |0\rangle_F$, we can show from (10) that the evolved state $\hat{U}(t)|\gamma\rangle_S |0\rangle_F$ is also an eigenstate of $\hat{c}$ with eigenvalue $\gamma(t) = \lambda^{-1} d\alpha(t)/dt$ for all times *t*. Therefore $|\gamma\rangle_S$ is an eigenstate of $\hat{c}$. Thus for a quantum system to produce a field in a coherent state as described above, the initial state of the source must be coherent in the sense of being in an eigenstate of $\hat{c}$. Because in this case $\langle \hat{a}^\dagger \hat{c} - \hat{c}^\dagger \hat{a} \rangle$ vanishes initially and $(\hat{a}^\dagger \hat{c} - \hat{c}^\dagger \hat{a})$ commutes with the Hamiltonian, $\alpha^*(t)\gamma(t)$ must be real at all times. Thus the argument of $\alpha(t)$ must be the same as the argument of $\gamma(t)$.

An important attribute of a coherent state $|\alpha\rangle$ is that it has a well-defined phase if reasonably intense. Its mean phase in a suitably chosen $2\pi$ window is the argument of $\alpha$ and its variance in the same window becomes quite small for a large mean photon number [19]. In the above discussion, we see that the mean phase of the coherent state produced by the source is the same as that of the source itself. That is, the phase information of the field comes from the source of the field, exactly as happens classically.

A particular example of a source such as that described above is a charged quantum mechanical harmonic oscillator. Here $\hat{c}$ will be an annihilation operator of the oscillator so the source state required to produce a coherent state of the field is a coherent state of the oscillator. Suppose, instead of representing a particular coherent state, the initial density operator of the oscillator can be written as

$$\hat{\rho}_S = \sum_{N=0}^{\infty} P_N |N\rangle_S {}_S\langle N| \tag{12}$$

where $|N\rangle_S$ are the energy eigenstates of the oscillator. Because of the energy conserving form of (9) the state $|0\rangle_F |N\rangle_S$ will evolve with time to a superposition

$$|C_N\rangle = \sum_{n=0}^{\infty} c_{n,N} |n\rangle_F |N-n\rangle_S \tag{13}$$

where $|n\rangle_F$ are the energy, or photon number eigenstates of the field. It follows straightforwardly that

$$\text{Tr}(|C_N\rangle\langle C_N|n-m\rangle_{F\ F}\langle n|) = \langle C_N|n-m\rangle_{F\ F}\langle n|C_N\rangle = 0 \qquad (14)$$

where $0 < m \leq n$. The reduced density operator $\hat{\rho}_F$ for the field at this time will be

$$\hat{\rho}_F = \text{Tr}_S\left(\sum_N P_N |C_N\rangle\langle C_N|\right). \qquad (15)$$

From (15) and (14) it follows that all off-diagonal elements of the reduced density operator representing the state of the field in the energy basis are zero. That is, there are no optical coherences. As shown in [20], the phase properties of the field are directly determined by the off-diagonal elements of $\hat{\rho}_F$:

$$\langle \cos(m\varphi)\rangle = \frac{1}{2}\sum_{n=0}^{\infty}\left(\langle n+m|\hat{\rho}_F|n\rangle + \langle n|\hat{\rho}_F|n+m\rangle\right) \qquad (16)$$

$$\langle \sin(m\varphi)\rangle = \frac{i}{2}\sum_{n=0}^{\infty}\left(\langle n|\hat{\rho}_F|n+m\rangle - \langle n+m|\hat{\rho}_F|n\rangle\right). \qquad (17)$$

Thus the vanishing of the off-diagonal elements means the phase probability distribution is uniform, or the phase is random, that is, a phase measurement of the field is equally likely to yield any value in a $2\pi$ range.

While physically it is not unreasonable that if we start off with random phase we end up with random phase, we must examine the nature of the randomness. Suppose in (12) we can say that the coefficients $P_N$ are *probabilities* that the associated oscillator energy states have been *prepared* by a preparation device, that is the oscillator is prepared in an energy state but we do not know which one. Then we can say that the field produced by the oscillator really has a purely random phase, that is, has no phase coherence at all. No extra knowledge of the preparation event would change the probability of a measurement yielding a particular value of phase. It would be a fiction to say that the field is in a coherent state with a well-defined but unknown phase. Suppose, on the other hand, that $P_N$ are *not* oscillator preparation probabilities but are equal, for example, to the corresponding coefficients in (2) and that $\hat{\rho}_S$ can be written in the form of (1) where the coefficients *are* preparation probabilities for coherent states of the oscillator. Then the field will be in a coherent state with a well-defined but unknown phase. With sufficient extra knowledge of the preparation event we could predict the outcome of a measurement to find this phase.

**5. Laser light**

In a laser there are mechanisms for exciting atoms, allowing the atoms to lase and for the light to escape from the cavity. The excitation is usually incoherent and the cavity losses are unlikely to generate coherence, so we shall examine the problem on a time scale such that the relevant part of the Hamiltonian is $\hat{H}_F + \hat{H}_A + \hat{H}_I$ where these terms are for the field, the system of atoms and the atom-field interaction. Such a time scale would be shorter than the characteristic pumping and cavity loss times. We assume that the atomic transitions that dominate the contribution have the same Bohr frequency $\omega$ as the field mode. The operator $\hat{c}$ in $\hat{H}_I$ given by (9) will be a linear combination of terms of the type $|g\rangle_i {}_i\langle e|$ where $|g\rangle_i$ and $|e\rangle_i$ are the lower and upper energy states of the *i*-th atom involved in the lasing transition. $\hat{H}_A$ and $\hat{H}_F$ are given by $\sum_i |e\rangle_i {}_i\langle e|\hbar\omega$ and $\hat{a}^\dagger \hat{a}\hbar\omega$ respectively.

Consider the case of incoherent excitation in which some mechanism excites the atoms to a state, which could be a pure multi-atom state but which will in general be a pure entangled state of the atoms and the exciting system, which we call $|A\rangle_S$. For incoherent excitation it is most probable that for this prepared state $\langle \hat{c} \rangle = 0$ and indeed $\langle \hat{C} \rangle = 0$ where $\hat{C}$ represents any linear combination of products of $\hat{c}$ and $\hat{c}^\dagger$ such that the number of factors $\hat{c}$ in each term is not equal to the number of factors $\hat{c}^\dagger$. Although any mixture of photon number states would suffice for our argument, to be specific, let the initial state of the field be the vacuum state $|0\rangle_F$. In the interaction representation the time displacement operator $\hat{U}$ will be $\exp(-i\hat{H}_I t)$ with $\hbar = 1$. After time $t$ the expectation value of $|n-m\rangle_{F\,F}\langle n|$ will be given by

$$_S\langle A|_F\langle 0|\hat{U}^\dagger|n-m\rangle_{F\,F}\langle n|\hat{U}|0\rangle_F|A\rangle_S \ . \tag{18}$$

For $m \neq 0$, expansion of $\exp(-i\hat{H}_I t)$ as a series shows that the non-zero terms in (18) will be of the form $_S\langle A|\hat{C}|A\rangle_S$. Thus the expectation value of $|n-m\rangle_{F\,F}\langle n|$ will vanish for $m \neq 0$ as it does in (14). It follows then that the reduced density matrix, $_F\langle n|\bar{\rho}_F|n'\rangle_F$ say, representing the state of the field in the energy basis is always diagonal and there are no optical coherences. The phase of the laser light is random for the prepared source state $|A\rangle_S$.

For a prepared source state for which $\langle \hat{C} \rangle \neq 0$ the laser light need not have a random phase. An example of this is an optical amplifier system in which the atoms are injected in prepared states that are controlled coherently excited superpositions such as $[|g\rangle_i + \exp(i\theta_i)|e\rangle_i]/\sqrt{2}$. In this case it is indeed possible to impress phase information onto the light [21, 22]. Then, if the angles $\theta_i$ are unknown but the value for each atom is correlated with those for other atoms, we can obtain light with at least a partially defined but unknown phase. The likelihood of preparation of such a source state by means of incoherent excitation of a very large number of atoms is, however, extremely small. Thus laser light will in general not have a definite but unknown phase. This is in accord with the conjecture of Mølmer [1].

In the above discussion, we have assumed that the field is initially in the vacuum state. It is not difficult to show that we would reach the same conclusion with the field initially in a mixture of photon number states. Again the initial lack of optical coherences is preserved. Sometimes, however, the action of the laser is thought of as amplifying an initial weak optical field that does have some non-uniform phase distribution but with an unknown mean. This might be caused, for example, by some accidental coherence in the excitation of some of the atoms associated with random fluctuations. The phase amplification properties of optical amplifiers have been studied in [22]. It is found that, for large amplification by a phase-insensitive amplifier, the phase variance of the amplified light is given by the input phase variance plus an extra term that is equal to the phase variance of a coherent state of the same intensity as the *initial* field. Thus the phase of the amplified field would be less well defined than that as the initial field and, as the initial field would be relatively weak, it must have a large phase variance. Thus this amplification process cannot give a very small phase variance commensurate with a strong coherent state.

Another physical picture is that of a coherent state with a well-defined phase angle that is performing a random walk around a circle [23]. This fits in with observed phase diffusion predicted by use of the complete Hamiltonian for the laser including the pumping and cavity loss terms [23]. In this picture, if a coherent state with a particular phase could be prepared and maintained in a time short compared with the diffusion time, then after a time long compared with the diffusion time it will still have a reasonably definite phase angle but this could be anywhere on the circle. That is, the laser would produce a coherent state with a definite but unknown phase. The unitary operator for a uniform shift $\Delta\varphi$ of the phase distribution is $\exp(-i\hat{a}^\dagger \hat{a}\Delta\varphi)$ [19]. This operator preserves the phase variance in a suitably chosen phase window but alters the mean phase. To generate a random walk would require an effective Hamiltonian of the form $\hat{a}^\dagger \hat{a} f(t)$ where $f(t)$ changes value randomly. For example $f(t)$ might change sign or not change sign at regular intervals $\Delta t$. It is not immediately obvious how such a Hamiltonian could be extracted from the full laser Hamiltonian. Thus, even if an initial coherent state could occur, it is not clear that the action of the phase diffusion process in broadening the phase distribution is to maintain the actual phase variance while randomly changing the mean.

To conclude this section, by examining the preparation of the light in a laser, we obtain agreement with Mølmer's conjecture. A coherent state is in general not one of the possible prepared states. Does this mean that the light is prepared in a definite but unknown photon number state as would be the case if the coefficients in (2) were preparation probabilities? This is unlikely. Even if the atoms are not entangled with the excitation system, in general the atoms and field in a laser cavity will be prepared in some entangled state. A reduced density operator for the field can be obtained by tracing over the atom states or, if necessary, over the states of the atoms and the excitation system. This reduced density operator can be used for predicting measurement probabilities only if the atom states, for example, are not measured, that is if the POM element for the measurement of the atoms is the unit operator on the state space of the atoms. If the atoms are entangled with the excitation system, we would include also the appropriate unit operator in the combined POM element. Thus in general the coefficients in a particular composition will depend on the POM element of a future measurement and thus cannot represent *a priori* preparation probabilities. An exceptional case would be if we had a source state, such as for a quantum oscillator, that did not become entangled with the field state. Then measuring the source state would not collapse the field state. We see, therefore, that although it can be used for predicting measurement probabilities provided the atom states are not measured, the reduced density operator is very much a fiction in itself. If the field itself could be prepared in a state given by, say, (2) where the coefficients are the genuine preparation probabilities, then the fiction in saying it is in a coherent state of unknown phase at least is undetectable by future measurements. On the other hand the fiction in using a reduced density operator for an atom-field system prepared in an entangled state is more serious in that it can be exposed by performing later measurements on the atoms.

**6. Possible observations**

In this section we examine two possible experiments that might appear at first glance capable of distinguishing a coherent state composition from other compositions.

The first is the experiment discussed by Mølmer [1, 3], Sanders *et al.* [24] and Cable *et al.* [25] which we treat here in a more general way to avoid using any fictional or particular states. The second involves deliberately disrupting the phase of light. This might be expected to affect a coherent state but not, for example, a photon number state that already has a random phase distribution.

*6.1. Interference between two lasers*

Classically, the concept of a well-defined phase is often associated with interference effects. While this gives a ready interpretation of the interference between two beams produced by splitting a single laser beam, it is well known that interference effects also occur if the original light that is split is, for example, in a photon number state. The interference is, in effect, an interference of the amplitudes for a photon to take the different paths in the interferometer. A more stringent test of the different possible compositions of laser light might therefore be the interference of light from two separate lasers [1]. Then, in an experiment occupying a time less that the diffusion time of each laser, there would be a well-defined phase difference in the coherent state picture that would lead to observable, and indeed observed, interferences [2]. In the complementary picture, in which we consider the laser fields to be in photon number states, there would be a well-defined number difference. This is the complement of phase difference and thus the laser fields would have a uniform phase difference distribution. Without a reasonably well-defined phase difference it might appear that interference is unlikely. Mølmer [1, 3] and Sanders *et al.* [24] have investigated this situation for the two cavity fields prepared in the identical photon number state and find, surprisingly, that interference effects do actually occur. Cable *et al*. [25] have extended the study and in particular have considered cases where the two cavities are in Poissonian and in thermal mixed states. As it is unlikely, however, that the two cavity fields are actually prepared in any unentangled field states such as these, we re-examine the problem here and present below a more general treatment in which we avoid using such fictional states or any particular cavity states at all.

Let us consider firstly the case of light leaking out of a *single* cavity and incident on a photodetector for a time much less than the phase-diffusion time. Let the internal cavity system have an initial density operator $\hat{\rho}$. This system can include the field inside the cavity, the atoms and any other essential system such as the excitation system if necessary. The probability for detecting a photon is the probability that the field outside the cavity, initially the vacuum state $|0\rangle_o$, becomes a one-photon state $|1\rangle_o$ where the subscript refers to the outside mode. The leaking mirror couples the inside field mode to the outside modes by means of an interaction Hamiltonian containing energy-conserving terms proportional to $\hat{a}_o^\dagger \hat{a}$ and $\hat{a}_o \hat{a}^\dagger$. Here a lack of a subscript implies an inside field mode. The combined initial state $\hat{\rho}|0\rangle_{o\,o}\langle 0|$ will evolve under this Hamiltonian a short time later to a sum of terms including a term $\varepsilon \hat{a} \hat{\rho} \hat{a}^\dagger |1\rangle_{o\,o}\langle 1|$, where $\varepsilon$ depends on the small time interval, and other terms involving states of the outside field orthogonal to $|1\rangle_o$. As we do not measure the state of the internal cavity system, the appropriate POM element is $|1\rangle_{o\,o}\langle 1|$, where we have not shown explicitly the unit operators acting on the atom, field or other state spaces inside the cavity. From (5) this gives the desired probability as being proportional to $\text{Tr}(\hat{a}\hat{\rho}\hat{a}^\dagger)$, where the trace is over the internal cavity system states, that is, proportional to the mean initial photon number $\bar{n}$ in the cavity. From the form of the term $\varepsilon \hat{a} \hat{\rho} \hat{a}^\dagger |1\rangle_o {}_o\langle 1|$, we see that if the field outside the cavity is found in state $|1\rangle_o {}_o\langle 1|$, then the state inside becomes proportional to $\hat{a} \hat{\rho} \hat{a}^\dagger$. That is, the detection of the first photon collapses the initial density operator inside the cavity from $\hat{\rho}$ to $\hat{a}\hat{\rho}\hat{a}^\dagger / \text{Tr}(\hat{a}\hat{\rho}\hat{a}^\dagger)$, after renormalisation. We might interpret this as one photon being destroyed in the cavity.

In the experiment in which we are interested, there are two similar cavities *a* and *b*, and the initial internal cavity systems have a combined density operator $\hat{\rho} = \hat{\rho}_a \hat{\rho}_b$. The fields leaving the cavities enter the input ports of a 50:50 beam splitter with photodetectors in its outputs. The POM element for detecting a photon in one photodetector and none in the other is $|1\rangle_1 |0\rangle_2 {}_2\langle 0|{}_1\langle 1|$ at the detectors. The subscripts 1 and 2 refer to the modes outside the two detectors. Applying a unitary backward-time evolution operator to the measurement POM element as discussed in Section 2, we find that this is transformed by the beam splitter into a POM element just outside the cavities given by $|f\rangle_o {}_o\langle f|$ where

$$|f\rangle_o = 2^{-1/2}\left(|1\rangle_{oa}|0\rangle_{ob} + e^{-i\gamma}|0\rangle_{oa}|1\rangle_{ob}\right) = 2^{-1/2}\left(\hat{a}_o^\dagger + e^{-i\gamma}\hat{b}_o^\dagger\right)|vac\rangle_o. \tag{19}$$

The subscript *oa* and *ob* refer to the modes outside the cavities *a* and *b*, $|vac\rangle_o = |0\rangle_{oa}|0\rangle_{ob}$ and the annihilation operators $\hat{b}$ and $\hat{b}_o$ act on field modes inside and outside the cavity *b* respectively. $\gamma$ is dependent on the path difference from the cavities to the beam splitter and on which detector registered the photocount. The interaction Hamiltonian connecting the inside to the outside modes for this case involves terms

proportional to $\hat{a}_o^\dagger \hat{a} + \hat{b}_o^\dagger \hat{b}$ and to $\hat{a}_o \hat{a}^\dagger + \hat{b}_o \hat{b}^\dagger$. The relevant term here is the first, which can be written as

$$\hat{a}_o^\dagger \hat{a} + \hat{b}_o^\dagger \hat{b} = \tfrac{1}{2}[(\hat{a}_o^\dagger + e^{-i\gamma}\hat{b}_o^\dagger)(\hat{a} + e^{i\gamma}\hat{b}) + (\hat{a}_o^\dagger - e^{-i\gamma}\hat{b}_o^\dagger)(\hat{a} - e^{i\gamma}\hat{b})] \; . \tag{20}$$

The term in the complete density operator that evolves in a short time from $\hat{\rho}|vac\rangle_{o\,o}\langle vac|$ arising from the first term on the right-hand side of (20) is given by

$$\hat{\rho}' \propto \left(\hat{a}_o^\dagger + e^{-i\gamma}\hat{b}_o^\dagger\right)|vac\rangle_o \left(\hat{a} + e^{i\gamma}\hat{b}\right)\hat{\rho}\left(\hat{a}^\dagger + e^{-i\gamma}\hat{b}^\dagger\right)_o \langle vac|\left(\hat{a}_o + e^{i\gamma}\hat{b}_o\right)$$

$$\propto |f\rangle_o \left(\hat{a} + e^{i\gamma}\hat{b}\right)\hat{\rho}\left(\hat{a}^\dagger + e^{-i\gamma}\hat{b}^\dagger\right)_o \langle f| \; . \tag{21}$$

The corresponding contribution arising from the second term in (20) involves a state orthogonal to $|f\rangle_o$, as do the contributions from all other terms to this order of smallness. Thus, with the POM element $|f\rangle_{o\,o}\langle f|$, the probability for detecting a photon in one photodetector and none in the other is proportional to

$$\text{Tr}[(\hat{a} + e^{i\gamma}\hat{b})\hat{\rho}(\hat{a}^\dagger + e^{-i\gamma}\hat{b}^\dagger)] = \bar{n}_a + \bar{n}_b + \left\langle \hat{a}^\dagger \hat{b} e^{i\gamma} + \hat{a}\hat{b}^\dagger e^{-i\gamma} \right\rangle \tag{22}$$

where Tr is the trace over the internal cavity system states of both cavities and $\bar{n}_a$ and $\bar{n}_b$ are the mean photon numbers for the initial cavity states. The detection of the first photon collapses the combined cavity atom-field state to

$$\hat{\rho}_1 = \frac{(\hat{a} + e^{i\gamma}\hat{b})\hat{\rho}(\hat{a}^\dagger + e^{-i\gamma}\hat{b}^\dagger)}{\text{Tr}[(\hat{a} + e^{i\gamma}\hat{b})\hat{\rho}(\hat{a}^\dagger + e^{-i\gamma}\hat{b}^\dagger)]} = \frac{(\hat{a} + e^{i\gamma}\hat{b})\hat{\rho}_a\hat{\rho}_b(\hat{a}^\dagger + e^{-i\gamma}\hat{b}^\dagger)}{\bar{n}_a + \bar{n}_b + \left\langle \hat{a}^\dagger \hat{b} e^{i\gamma} + \hat{a}\hat{b}^\dagger e^{-i\gamma} \right\rangle} \; . \tag{23}$$

after renormalisation.

To see how this collapse, or state reduction, has altered the properties of the light we examine the phase difference probability distribution $P(\Delta)$. This distribution is $2\pi$-periodic and thus can be written as a Fourier series

$$P(\Delta) = \frac{1}{2\pi} \sum_{p=-\infty}^{\infty} \exp(ip\Delta)\langle\exp(-ip\Delta)\rangle \ . \tag{24}$$

It has been shown in reference [26] that for physical states $\langle\exp(im\varphi)\rangle$ for canonical phase, as found for example by the limiting procedure in reference [19], is equal to $\langle\hat{E}^m\rangle$ where $\hat{E}$ is $\sum_n |n\rangle\langle n+1|$, the non-unitary Susskind-Glogower operator [27], and $m \geq 0$. Also $\langle\exp(-im\varphi)\rangle = \langle(\hat{E}^\dagger)^m\rangle$. Similarly for the canonical phase difference we can show for physical states that $\langle\exp(im\Delta)\rangle = \langle\hat{E}_a^m(\hat{E}_b^\dagger)^m\rangle$, which is just $\text{Tr}[\hat{\rho}\hat{E}_a^m(\hat{E}_b^\dagger)^m]$, for $m \geq 0$. The complex conjugate gives $\langle\exp(-im\Delta)\rangle$.

We consider the case where the initial internal cavity states are such that the expectation values of $|n'\rangle_a{}_a\langle n|$ and $|n'\rangle_b{}_b\langle n|$ are zero for photon numbers $n \neq n'$, that is where the optical coherences vanish. Then, from the above, we can show that the only non-zero term in (24) will be for $p = 0$. Thus $P(\Delta) = 1/(2\pi)$, which is a uniform, or random, distribution. In addition to the internal cavity states discussed in the previous section, this case is also applicable to density operators given by (1) and (2). It is not difficult to show that in this case the expectation values of $\hat{a}$ and $\hat{b}$ are also zero, so the collapsed state after the first photon detection becomes from (23) just

$$\hat{\rho}_1 = \frac{(\hat{a} + e^{i\gamma}\hat{b})\hat{\rho}_a\hat{\rho}_b(\hat{a}^\dagger + e^{-i\gamma}\hat{b}^\dagger)}{\bar{n}_a + \bar{n}_b} \ . \tag{25}$$

From (25) it is straightforward to find, remembering that the expectation values of $|n'\rangle_a{}_a\langle n|$ and $|n'\rangle_b{}_b\langle n|$ vanish for $n \neq n'$, that only the terms in (24) with $p = 0, \pm 1$ are

non zero. Using $\hat{a} = \hat{E}_a \hat{N}_a^{1/2}$, where $\hat{N}_a$ is the photon number operator, and associated relations we obtain from (25), as shown in Appendix A,

$$P(\Delta) = \frac{1}{2\pi} + \frac{1}{\pi} \frac{\overline{n_a^{1/2} n_b^{1/2}}}{\overline{n}_a + \overline{n}_b} \cos(\Delta - \gamma) \quad . \tag{26}$$

This subsequent narrowing shown by (26) from the uniform distribution following the detection of the first photon increases the chances of the second photon being detected at the same detector as the first, which would lead to a further narrowing and so on. If the mean initial photon number in one cavity is much greater than in the other, the second term in (26) is very small and the distribution remains uniform. The narrowing effect is most pronounced for initial internal cavity states with narrow photon number distributions and with $\overline{n}_a = \overline{n}_b$ for which (5.4) reduces to $[1 + \cos(\Delta - \gamma)]/(2\pi)$. Then the phase difference variance, in a phase window chosen such that the peak of the distribution is in the centre, is reduced substantially from the random value of $\pi^2/3$ to $(\pi^2/3) - 2$. It is interesting that the interference effects depend on the narrowness of the *number* state distribution.

The probabilities for the detection of the second photon are also interesting. The probability $P_{11}$ that the same detector as detected the first photon will also detect the second photon can be found by calculating the left-hand side of (22) with $\hat{\rho}$ replaced by $\hat{\rho}_1$ given by (23). To find the probability $P_{12}$ that the other detector detects the second photon, we first change $\gamma$ in the left-hand side of (22) to $\gamma + \pi$. In the case where both internal cavity states are pure coherent states of light we find from (23) that $\hat{\rho}_1 = \hat{\rho}$. Thus the detection of the first photon does not affect the probability of where the second photon will be detected. In the case where the optical coherences vanish, on the other hand, we find in Appendix A, using (25), that the ratio of the probabilities is

$$\frac{P_{12}}{P_{11}} = \frac{\overline{n_a^2} + \overline{n_b^2} - \overline{n}_a - \overline{n}_b}{\overline{n_a^2} + \overline{n_b^2} - \overline{n}_a - \overline{n}_b + 4\overline{n}_a \overline{n}_b}. \tag{27}$$

When the initial mean photon number for one cavity is much greater than for the other such that, for example, $4\overline{n}_a \overline{n}_b << \overline{n_a^2}$, this ratio reduces to unity and it is equally likely for the second photon to be detected in either detector. This is in accord with the phase difference distribution remaining uniform. Some extreme cases for narrow initial number state distributions with equal initial mean photon numbers are as follows. If these mean photon numbers are much greater than unity, the ratio reduces to 1/3, that is, it is three times more likely for the second photon to be detected by the detector that detected the first photon than by the other detector. This agrees with the figure quoted by Mølmer for the case where both cavities are in the identical pure number state [3]. If on the other hand there is initially only one photon in each cavity, the effect is even *more* pronounced with the ratio reducing to zero. In this case the second photon *must* be detected by the detector that detected the first. This effect has, in fact, been verified experimentally by Hong, Ou and Mandel [28]. When two photons are incident on a 50:50 beam splitter, they must both be detected by the same detector.

Mølmer [1, 3] has numerically simulated interference graphs for the special case where both cavity fields are in the identical pure number state and this case has also been examined by Sanders *et al*. [24]. Cable *et al*. have extended the study to include Poissonian and thermal mixed states [25]. Mølmer's graphs are similar to those expected from the coherent state picture and which have been observed experimentally [2]. The

27phase narrowing result derived above and the enhanced probability for the second photon being detected by the same detector as the first photon depend only on the internal cavity density operators $\hat{\rho}_a$ and $\hat{\rho}_b$ being such that the expectation values of $|n'\rangle_a{}_a\langle n|$ and $|n'\rangle_b{}_b\langle n|$ vanish for photon numbers $n \neq n'$. Thus the same interference effects will follow irrespective of whether or not we can factorise the internal cavity states into separate atom and field states. These effects are also independent of how we might decompose the resulting individual field density operators, which could be as number states, coherent states or neither. This result is based on the concept of the internal cavity state being entangled with the field state outside the cavity, which allows the internal state to be collapsed by measurement of the outside state. We have seen, on the other hand, that pure internal coherent field states, such as produced by oscillators inside the cavities, are not collapsed by the photon detection. Essentially this is because there is no such entanglement with the outside states. It may seem a little puzzling, therefore, that for a mixture of coherent states the probability of the second measurement can be affected by the outcome of the first. For the coherent state case, the mechanism is retrodiction rather than state collapse. The first measurement provides information that changes the classical probabilities from the *a priori* probabilities associated with the individual coherent states in the initial mixed density operator. In Appendix B we give a formal retrodictive analysis of this case and show that the *a posteriori* density operator is identical to $\hat{\rho}_1$ given by (25).

We see from the above that, beginning with two sources that are not expected to show interference because of their uniform phase difference distribution, detecting a photon by a method which makes it impossible to tell from which source the photon originated can produce a phase difference. This allows interference effects for subsequently emitted light. It is interesting to note this concept is not new and has indeed been experimentally verified by Ghosh and Mandel [29]. By means of a parametric down-conversion process they used sources in number states with $n_a = n_b = 1$. For a description of this experiment in terms of the phase difference induced by collapse after detection of the first photon see reference [30]. There it was found that the phase difference variance was reduced from an initial $\pi^2/3$ to $(\pi^2/3) - 2$ in an appropriately chosen $2\pi$ range, precisely in accord with that obtained from (26) with $\bar{n}_a = \bar{n}_b$. We note that although photon detection results in entanglement of the cavity fields if the initial states are photon number states, entanglement is not necessary for a relative phase to develop. Cable *et al.* [25] have shown that the cavity states remain separable if they are initially Poissonian or thermal mixed states.

*6.2. Phase disruption of laser light*

An alternative way of experimentally distinguishing between a coherent state of unknown phase and a mixture of number states would appear to be as follows. While coherent states have a narrow phase distribution, states such as number states or mixtures of number states have a uniform distribution. Thus deliberately *disrupting* the phase of the light from a laser might be expected to affect a coherent state of unknown phase but not affect a number state. Essentially this is because a number state is an eigenstate of $\hat{a}^\dagger \hat{a}$, which is the generator of the phase shift. Can we detect this difference experimentally, for example by studying the effect of the disruption on the excitation of a two-level atom? If we restrict ourselves to a time scale much shorter than the diffusion time, then a coherent state of unknown phase will maintain this phase if used as a π-pulse for exciting a two-level atom from the ground state $|g\rangle$ to the excited state $|e\rangle$. We may think of the action of a strong coherent state as being similar to that of a classical field, which causes the Bloch vector to precess steadily from $|g\rangle$ to $|e\rangle$ during the pulse time. The probability of the atom then being found in $|e\rangle$ is then unity, irrespective of the actual unknown phase of the coherent state, provided the same phase is maintained throughout the pulse. It is well known that a field in a photon number state, or with a narrow number state distribution, can also excite the atom from $|g\rangle$ to $|e\rangle$ in an appropriate time [18]. Such a field does not have a particular phase. Now consider the case in which we deliberately disrupt the phase of the pulse during the pulse. We could do this, for example, by applying a phase shift halfway through the pulse. If the field is in a coherent state during the pulse, we should reduce the probability of the atom being found in $|e\rangle$ to less than unity. Indeed, if this were a phase shift of π, it should reduce the probability of the atom being found in $|e\rangle$ to zero. On the other hand, applying a phase shift to a state, such as a number state, with a uniform phase distribution does not alter the state. Thus we might expect that the probability of finding the atom in $|e\rangle$ is still unity. In such a case we would have a way of experimentally distinguishing the two pictures. We now examine the situation in detail.

We choose the frequency of the light to equal the Bohr frequency of the atom and let the pulse be in a number state $|n\rangle$. The interaction Hamiltonian is of the form [18]

$$\hat{H}_I = i\lambda \left( \hat{a}^\dagger |g\rangle\langle e| - \hat{a}|e\rangle\langle g| \right) . \tag{28}$$

An initial state of $|g\rangle|n\rangle$ will, because of the energy conserving form of (28), evolve to a superposition $c_g(t)|g\rangle|n\rangle + c_e(t)|e\rangle|n-1\rangle$. Using this state in the Schrödinger equation with Hamiltonian (28) yields two simple coupled equations which are easily solved to give [18]

$$c_g(t) = \cos(\sqrt{n}\lambda t) \tag{29}$$

$$c_e(t) = -\sin(\sqrt{n}\lambda t). \tag{30}$$

The time required for a π-pulse is thus $t_\pi = \pi/(2\sqrt{n}\lambda)$. Let us now apply a π phase shift to the field at time $t_\pi/2$ and then allow the field to interact with the atom for another period of $t_\pi/2$. At $t_\pi/2$ state $|g\rangle|n\rangle$ will have evolved to $(|g\rangle|n\rangle - |e\rangle|n-1\rangle)/\sqrt{2}$. The unitary operator giving a π phase shift is $\hat{U}(\pi) = \exp(-i\hat{a}^\dagger \hat{a}\pi)$. Applying this to the evolved state yields the state $(-1)^n(|g\rangle|n\rangle + |e\rangle|n-1\rangle)/\sqrt{2}$. We can show similarly to the above that a state $|e\rangle|n-1\rangle$ will evolve in time $t$ to $\cos(\sqrt{n}\lambda t)|e\rangle|n-1\rangle + \sin(\sqrt{n}\lambda t)|g\rangle|n\rangle$. Thus after another period of $t_\pi/2$, the final state will be $(-1)^n|g\rangle|n\rangle$. Thus the atom will be left in its ground state just as for the coherent state case. We see that it is the *entanglement* between the atom and the field induced by the first $\pi/2$ pulse that allows the atom-field system to be affected by the phase disturbing pulse. There is a non-zero amplitude for the field to be found in $|n\rangle$ after the first $\pi/2$ pulse and a non-zero amplitude for the field to be found in $|n-1\rangle$. A superposition of $|n\rangle$ and $|n-1\rangle$ suffers a phase shift under the action of $\hat{U}(\pi)$.

Of course, just as with the interference between two lasers, it is not merely a coincidence that the results of this experiment are in accord with the preferred ensemble fallacy. We have shown that for the first experiment that the probability of a particular detector registering two consecutive photocounts depends only on the initial density operator and not its particular composition. In the second experiment the three unitary time-evolution operators involved can be combined mathematically into a single unitary operator $\exp(-i\hat{H}_I t_\pi / 2)\hat{U}(\pi)\exp(-i\hat{H}_I t_\pi / 2)$, which can be reduced simply to $\hat{U}(\pi)$. This leaves the atom in the ground state regardless of the initial field state.

**Conclusion**

We have examined the controversial conjecture of Mølmer [1] that optical coherences in laser light may well be just a convenient fiction. Mathematically we can write the reduced density operator for laser light as a mixed state representing, for example, a coherent state of unknown phase or a photon number state of unknown number. If the coherent state description is a fiction then, as pointed out by Rudolph and Sanders [7], continuous variable teleportation as implemented by experiments so far may also be a fiction. On the other hand the preferred ensemble fallacy implies that measurements made on the light cannot distinguish between different descriptions such as those above. On this basis it might be argued that no particular description is any more real than any other and we can thus choose the most convenient. van Enk and Fuchs [8] disagree with the claims of Rudolph and Sanders and propose that coherent states do play a privileged role in the description of laser light. Wiseman [11] also defends continuous variable teleportation and argues that a laser beam is used as a clock and it is as good a clock as any other. We do not discuss these arguments here.

In this paper we have examined the question of deciding between what is real and what is fictional. Although different compositions of the density operator cannot be distinguished by future measurement events, they are not equivalent when past preparation events are considered. When the coefficients of components of a mixed state represent actual *a priori* preparation probabilities these components are not fictional. For example, suppose Alice prepares a spin-half atom with a density operator proportional to the unit operator by selecting state $|+z\rangle$ or $|-z\rangle$ with equal *a priori* probability. Such a mixed state is indistinguishable by future measurement information from an equal mixture of $|+y\rangle\langle+y|$ and $|-y\rangle\langle-y|$. However, with suitable information about the orientation of Alice's preparation device we can say, for example, that the probability that the atom has been prepared in state $|+z\rangle$ is 1/2 and the probability that the atom has been prepared in state $|+y\rangle$ is zero. Thus, although it may be convenient for predictive purposes to decompose the density operator in terms of the *y*-states, it would be a fiction to say that these were the states that were actually prepared.

By examining the preparation procedure for laser light we conclude, in agreement with Mølmer that the optical coherences corresponding to the composition (1) are a fiction. The composition in terms of photon number states, however, also appears to be a fiction. The actual state inside the laser cavity is most likely an entangled state in involving the field, the atoms and possibly the excitation mechanism. A reduced density operator for the field can be obtained by tracing over the appropriate part of the system. This would mean, however, that the coefficients in the resulting mixture would depend on a future event, the non-measurement of the part of the system over which the trace was taken, and thus cannot represent *a priori* preparation probabilities. Consequently we have generalised Mølmer's demonstration that two cavities containing fields in the same number state give the same interference effects as two coherent states. We have shown that the collapse of the entangled internal cavity state induced by a measurement of the external field gives the same result as the change of probabilities in a coherent state mixture due to retrodiction on the basis of the measurement outcome.

As a potential means of discriminating between coherent states and number states in addition to that considered by Mølmer, we have studied the effect of deliberately disrupting the phase of the light outside the laser by appropriate phase shifting. One might expect that this would destroy the phase coherence of a coherent state but leave unaffected a photon number state, which already has a random phase. We find, however, that by using a two-level atom as a detector we still cannot distinguish between the disrupted coherent state and the disrupted number state.

In conclusion, we must agree that although describing laser light as a coherent state is a fiction that cannot be revealed by future measurements of the light, it is nevertheless a fiction. Using a reduced density operator to describe the light, however, is also a fiction. It is convenient and legitimate to use such descriptions to predict the outcomes of future measurements, provided the fictional nature of the reduced density operator is not exposed by measuring the atoms. It is important, however, not to confuse fiction with reality when inferring from experimental results whether or not processes that rely on the coherent state assumption have actually occurred, for example, whether continuous variable teleportation has been implemented or whether a particular state, such as a squeezed state, has really been prepared.

**Acknowledgments**

DTP acknowledges financial support from the Australian Research Council and discussions with Barry Sanders. JJ thanks the U.K. Engineering and Physical Sciences Research Council for financial assistance.

**Appendix A**

*Derivation of equations (26) and (27)*

The initial expectation values of $|n'\rangle_a{}_a\langle n|$ and $|n'\rangle_b{}_b\langle n|$ are zero for photon numbers $n \neq n'$. Because of this, expressions such as $\text{Tr}_a(\hat{\rho}_a \hat{a})$ and $\text{Tr}_a(\hat{\rho}_a \hat{E}_a)$ vanish because they contain lowering operators such as $|n-1\rangle_a{}_a\langle n|$. Likewise expressions such as $\text{Tr}_a(\hat{\rho}_a \hat{E}_a \hat{a})$ will also vanish. On the other hand, expressions such as $\text{Tr}_a(\hat{\rho}_a \hat{a}^\dagger \hat{E}_a)$ will not vanish because the raising action of $\hat{a}^\dagger$ counteracts the lowering action of $\hat{E}_a$.

We wish to find $\text{Tr}_{ab}[\hat{\rho}_1 \hat{E}_a^m (\hat{E}_b^\dagger)^m]$ for $m > 0$, as it is clearly unity for $m = 0$. Substitution for $\hat{\rho}_1$ from (23) yields four terms. For $m = 1$ the only term that does not vanish has a numerator

$$e^{i\gamma}\text{Tr}(\hat{b}\hat{\rho}_a\hat{\rho}_b\hat{a}^\dagger \hat{E}_a \hat{E}_b^\dagger) = e^{i\gamma}\text{Tr}_b(\hat{b}\hat{\rho}_b\hat{E}_b^\dagger)\text{Tr}_a(\hat{\rho}_a\hat{a}^\dagger \hat{E}_a). \tag{A.1}$$

This is the only term that has a balance between raising and lowering operators for both cavity states. To calculate this we use the relation $\hat{E}_a^\dagger \hat{a} = \hat{N}_a^{1/2} = \hat{a}^\dagger \hat{E}_a$, which can be easily checked by allowing each expression to act on any number state $|n\rangle_a$. After using the cyclic property of the trace to write the first factor on the right of (A.1) as

$e^{i\gamma}\text{Tr}_b(\hat{E}_b^\dagger \hat{b}\hat{\rho}_b)$, we can use a similar simplifying relation to write this as $e^{i\gamma}\text{Tr}_b(\hat{N}_b^{1/2}\hat{\rho}_b)$.

Expression (A.1) then reduces to $e^{i\gamma}\overline{n_b^{1/2}}\,\overline{n_a^{1/2}}$.

For $m \geq 2$ there are insufficient factors of $\hat{a}^\dagger$, for example, in any term to balance $\hat{E}_a^m$. We thus obtain no contribution to $P(\Delta)$ in (24) from such terms. The contribution from $m = -1$ is just the complex conjugate of (A.1), that is $e^{-i\gamma}\overline{n_b^{1/2}}\,\overline{n_a^{1/2}}$. Combining the contributions for $m = 0$ and $m = \pm 1$, including the denominator in (25) yields (26).

To derive (27) we note that

$$P_{11} \propto \text{Tr}[(\hat{a}+e^{i\gamma}\hat{b})\hat{\rho}_1(\hat{a}^\dagger+e^{-i\gamma}\hat{b}^\dagger)] \ . \tag{A.2}$$

When we substitute for $\hat{\rho}_1$ from (23) we obtain 16 terms. Because the initial expectation values of $|n'\rangle_a{}_a\langle n|$ and $|n'\rangle_b{}_b\langle n|$ are zero for photon numbers $n \neq n'$, the only non-zero terms will contain a lowering operator to counteract the effect of a raising operator. That is, it must contain an $\hat{a}^\dagger$ for every $\hat{a}$ and a $\hat{b}^\dagger$ for every $\hat{b}$. Thus there are only six non-zero terms, which are proportional to $\text{Tr}(\hat{a}\hat{a}\hat{\rho}_a\hat{\rho}_b\hat{a}^\dagger\hat{a}^\dagger)$, $\text{Tr}(\hat{b}\hat{b}\hat{\rho}_a\hat{\rho}_b\hat{b}^\dagger\hat{b}^\dagger)$, $\text{Tr}(\hat{a}\hat{b}\hat{\rho}_a\hat{\rho}_b\hat{a}^\dagger\hat{b}^\dagger)$, $\text{Tr}(\hat{a}\hat{b}\hat{\rho}_a\hat{\rho}_b\hat{b}^\dagger\hat{a}^\dagger)$, $\text{Tr}(\hat{b}\hat{a}\hat{\rho}_a\hat{\rho}_b\hat{a}^\dagger\hat{b}^\dagger)$ and $\text{Tr}(\hat{b}\hat{a}\hat{\rho}_a\hat{\rho}_b\hat{b}^\dagger\hat{a}^\dagger)$. Each of the last four terms is equal to

$$\text{Tr}_a(\hat{a}^\dagger\hat{a}\hat{\rho}_a)\text{Tr}_b(\hat{b}^\dagger\hat{b}\hat{\rho}_b) = \overline{n}_a\overline{n}_b \ . \tag{A.3}$$

The first term can be written as $\text{Tr}(\hat{a}^\dagger\hat{a}^\dagger\hat{a}\hat{a}\hat{\rho}_a\hat{\rho}_b)$, which becomes

$$\text{Tr}[\hat{a}^\dagger(\hat{a}\hat{a}^\dagger-1)\hat{a}\hat{\rho}_a\hat{\rho}_b] = \overline{n_a^2} - \overline{n}_a. \tag{A.4}$$

The second term gives a corresponding result. Thus

$$P_{11} \propto \overline{n_a^2} + \overline{n_b^2} - \overline{n}_a - \overline{n}_b + 4\overline{n}_a\overline{n}_b \ . \tag{A.5}$$

To find $P_{12}$, we change $\gamma$ to $\gamma + \pi$ in (22), but not in (23). This again leads to six non-zero terms. The first two are the same as before. The last four have the same magnitudes as before but two of them have their signs reversed. Thus the total contribution from the last four terms is zero and we obtain

$$P_{12} \propto \overline{n_a^2} + \overline{n_b^2} - \overline{n}_a - \overline{n}_b .\qquad (A.6)$$

**Appendix B**

*Coherent states and retrodiction*

Consider a case where Alice can actually prepare a large number of coherent states with equal amplitude but different phases in a cavity, for example by use of a quantum oscillator and a phase shifter. Let $P_a(i)$ be the *a priori* probability that the state $|\alpha_i\rangle_a$ is prepared. If the state $|\alpha_i\rangle_a$ is prepared then after a short time the combined field inside and outside the cavity will evolve to $|\alpha_i\rangle_a |\varepsilon\alpha_i\rangle_{oa}$ where $\varepsilon$ is very small and, to this order of approximation, we write $\alpha_i$ instead of $\alpha_i(1 - \varepsilon^2/2)$ for the inside field. Then the probability that the state $|\varepsilon\alpha_i\rangle_{oa}$ outside the cavity is prepared is also $P_a(i)$. Because of the lack of entanglement, a measurement made on the field outside the cavity will not collapse the inside field state. As the inside and outside field states are correlated, however, the result of the measurement will give some information about the inside field that we can use to modify the initial probabilities. This is essentially a problem in retrodiction. The preparation device operator associated with the preparation of $|\varepsilon\alpha_i\rangle_{oa}$ is $\hat{\Lambda}_a(i) = P_a(i) |\varepsilon\alpha_i\rangle_{oa\ oa}\langle\varepsilon\alpha_i|$. We now apply the formal theory of retrodiction to the two-cavity problem.

We firstly apply retrodiction to the state outside the cavities *a* and *b* based on the measurement outcome that one photon is detected by one photodetector and none by the other. The POM element for this detection event is $\hat{\Pi}_j = |f\rangle_{o\,o}\langle f|$ with $|f\rangle_o$ given by (19). $P_b(k)$ is the preparation probability of $|\beta_k\rangle_{b\,b}\langle\beta_k|$ in cavity *b*. The combined preparation device operator for the combined field state $|\varepsilon\alpha_i\rangle_{oa\,oa}\langle\varepsilon\alpha_i| \otimes |\varepsilon\beta_j\rangle_{oa\,oa}\langle\varepsilon\beta_j|$ outside both cavities is given by

$$\hat{\Lambda}_a(i)\hat{\Lambda}_b(k) = P_a(i)P_b(k)|\varepsilon\alpha_i\rangle_{oa\,oa}\langle\varepsilon\alpha_i| \otimes |\varepsilon\beta_k\rangle_{ob\,ob}\langle\varepsilon\beta_k|. \tag{B.1}$$

From (6) with $\hat{\Lambda}_a(i)\hat{\Lambda}_b(i)$ in place of $\hat{\Lambda}_i$ we find that the probability that the field state outside the cavities was $|\varepsilon\alpha_i\rangle_{oa\,oa}\langle\varepsilon\alpha_i| \otimes |\varepsilon\beta_j\rangle_{ob\,ob}\langle\varepsilon\beta_j|$ is proportional to

$$P_a(i)P_b(k)\,_o\langle vac|(\hat{a}_o + e^{i\gamma}\hat{b}_o)|\varepsilon\alpha_i\rangle_{oa\,oa}\langle\varepsilon\alpha_i| \otimes |\varepsilon\beta_k\rangle_{ob\,ob}\langle\varepsilon\beta_k|(\hat{a}_o^\dagger + e^{-i\gamma}\hat{b}_o^\dagger)|vac\rangle_o$$

$$\propto P_a(i)P_b(k)\left|\alpha_i + e^{i\gamma}\beta_k\right|^2. \tag{B.2}$$

This must also be proportional to the probability that the field prepared inside the cavities was $|\alpha_i\rangle_{a\,a}\langle\alpha_i| \otimes |\beta_j\rangle_{o\,o}\langle\beta_j|$. Thus the retrodicted density operator of the field inside the cavities is proportional to

$$\sum_{i,k} P_a(i)P_b(k)\left|\alpha_i + e^{i\gamma}\beta_k\right|^2 |\alpha_i\rangle_{a\,a}\langle\alpha_i| \otimes |\beta_k\rangle_{b\,b}\langle\beta_k| = (\hat{a} + e^{i\gamma}\hat{b})\hat{\rho}(\hat{a}^\dagger + e^{-i\gamma}\hat{b}^\dagger) \tag{B.3}$$

where

$$\hat{\rho} = \sum_{i,k} P_a(i)P_b(k)|\alpha_i\rangle_{a\,a}\langle\alpha_i| \otimes |\beta_k\rangle_{b\,b}\langle\beta_k| \tag{B.4}$$

is the *a priori* density operator assigned to the state of the cavities before the measurement. The retrodicted, or *a posteriori*, density operator (B.3) can be normalised by division by its trace. Comparison with (25) shows that the retrodicted state, that is the state that has been modified on the basis of measurement information, is precisely the same as the measurement-collapsed state with density operator $\hat{\rho}_1$ calculated on the basis of entanglement between the states inside and outside the cavity. Of course $\hat{\rho}_1$ in (25) refers to the internal cavity systems whereas (B.4) refers to the internal fields. In this case, we have quantum oscillators instead of atoms and the density operator for the internal cavity system in each cavity can be factorised. This allows us to multiply (B.4) by the density operators for the oscillators and complete the correspondence with (25).